\documentclass[runningheds]{llncs}

\usepackage{amsfonts}
\usepackage{amsbsy,amssymb,amsmath}
\usepackage{todonotes}
\usepackage{latexsym}
\usepackage{url}
\usepackage{listings}
\usepackage{alltt}
\usepackage{graphicx}
\usepackage{color}
\usepackage{colortbl}
\usepackage{hyperref}
\usepackage{tikz}
\usetikzlibrary{arrows,positioning,calc,shapes.geometric}
\usepackage{caption}
\usepackage{subcaption}

\newcommand{\RR}{\mathbb{R}}
\newcommand{\QQ}{\mathbb{Q}}
\newcommand{\ZZ}{\mathbb{Z}}
\newcommand{\myRightarrow}{\; \Rightarrow \;}
\newcommand{\fn}[1]{\mathtt{#1}} 
\newcommand{\nroot}{\fn{root}}
\newcommand{\abs}{\fn{abs}}

\pdfpageattr {/Group << /S /Transparency /I true /CS /DeviceRGB>>}

\hypersetup{
    colorlinks=true,
    citecolor=blue,
    linkcolor=blue,
    urlcolor=blue
}

\tikzset{
    >=stealth',
    module/.style={
           rectangle,
           draw=black, very thick,
           text width=7em,
           minimum height=2em,
           text centered},
    interface/.style={
           ->,
           thick,
           shorten <=2pt,
           shorten >=2pt,}
}

\begin{document}

\title{A heuristic prover for real inequalities}


\author{Jeremy Avigad\inst{1} \and Robert Y. Lewis\inst{1} \and Cody Roux\inst{2}}

\institute{Carnegie Mellon University, Pittsburgh, PA 15213, USA \and Draper Laboratories, Cambridge, MA 02139, USA}

\maketitle

\begin{abstract}
We describe a general method for verifying inequalities between real-valued expressions, especially the kinds of straightforward inferences that arise in interactive theorem proving. In contrast to approaches that aim to be complete with respect to a particular language or class of formulas, our method establishes claims that require heterogeneous forms of reasoning, relying on a Nelson-Oppen-style architecture in which special-purpose modules collaborate and share information. The framework is thus modular and extensible. A prototype implementation shows that the method works well on a variety of examples, and complements techniques that are used by contemporary interactive provers.
\end{abstract}

\section{Introduction}
\label{section:introduction}

Comparing measurements is fundamental to the sciences, and so it is not surprising that ordering, bounding, and optimizing real-valued expressions is central to mathematics. A host of computational methods have been developed to support such reasoning, using symbolic or numeric methods, or both. For example, there are well-developed methods of determining the satisfiability or unsatisfiability of linear inequalities \cite{pugh:92} \cite{schrijver:86}, polynomial inequalities \cite{basu:et:al:03}, nonlinear inequalities involving functions that can be approximated numerically \cite{gao:et:al:12} \cite{moore:et:al:09}, and inequalities involving convex functions \cite{boyd:vandenberghe:04}. The ``satisfiability modulo theories'' framework \cite{barrett:et:al:08} \cite{nelson:oppen:79} provides one way of integrating such methods with ordinary logical reasoning and proof search; integration with resolution theorem proving methods has also been explored \cite{akbarpour:paulson:08} \cite{prevosto:waldmann:06}. Interactive theorem provers like Isabelle \cite{nipkow:et:al:02} and HOL Light \cite{harrison:07c} now incorporate various such methods, either constructing correctness proofs along the way, or reconstructing them from appropriate certificates. (For a small sample, see \cite{blanchette:et:al:11} \cite{chaieb:nipkow:08} \cite{harrison:07b} \cite{mclaughlin:harrison:05}.)
Such systems provide powerful tools to support interactive theorem proving. But, frustratingly, they often fail when it comes to fairly routine calculations, leaving users to carry out explicit calculations painstakingly by hand. Consider, for example, the following valid implication:
\[
 0 < x < y, \; u < v \myRightarrow 2 u + \fn{exp}(1 + x + x^4) < 2 v + \fn{exp}(1 + y + y^4)
\]
The inference is not contained in linear arithmetic or even the theory of real-closed fields. The inference is tight, so symbolic or numeric approximations to the exponential function are of no use. Backchaining using monotonicity properties of addition, multiplication, and exponentiation might suggest reducing the goal to subgoals $2 u < 2 v$ and $\fn{exp}(1 + x + x^4) < \fn{exp}(1 + y + y^4)$, but this introduces some unsettling nondeterminism. After all, one could just as well reduce the goal to
\begin{itemize}
\item $2 u < \fn{exp}(1 + y + y^4)$ and $\fn{exp}(1 + x + x^4) < 2 v$, or
\item $2 u + \fn{exp}(1 + x + x^4) < 2 v$ and $0 < \fn{exp}(1 + y + y^4)$, or even
\item $ 2 u < 2 v + 7$ and $\fn{exp}(1 + x + x^4) < \fn{exp}(1 + y + y^4) - 7$.
\end{itemize}
And yet, the inference is entirely straightforward. With the hypothesis $u < v$ in mind, you probably noticed right away that the terms $2u$ and $2 v$ can be compared; similarly, the comparison between $x$ and $y$ leads to comparisons between $x^4$ and $y^4$, then $1 + x + x^4$ and $1 + y + y^4$, and so on. 

The method we propose is based on such heuristically guided forward reasoning, using properties of addition, multiplication, and the function symbols involved. As is common for resolution theorem proving, we try to establish the theorem above by negating the conclusion and deriving a contradiction. We then proceed as follows:
\begin{itemize}
 \item Put all terms involved into a canonical normal form. This enables us to recognize terms that are the same up to a scalar multiple, and up to associativity and commutativity of addition and multiplication.
 \item Iteratively call specialized modules to learn new comparisons between subterms, and add these new comparisons to a common ``blackboard'' structure, which can be accessed by all modules.
\end{itemize}
The theorem is verified when any given module derives a contradiction using this common information. The procedure fails when none of the modules can learn anything new. We will see in Section~\ref{section:examples} that the method is far from complete, and may not even terminate. On the other hand, it is flexible and extensible, and easily verifies a number of inferences that are not obtained using more principled methods. As a result, it provides a useful complement to more conventional approaches.

We have designed and implemented modules to learn comparisons from the additive and multiplicative structure of terms, a module to instantiate axioms involving arbitrary functions symbols, and special-purpose modules for common functions like min, max, absolute value, exp, and log. The additive and multiplicative modules have two different implementations, with different characteristic strengths and weaknesses. The first uses a natural but naive Fourier-Motzkin elimination, and the second uses more refined geometric techniques. Our prototype implementation, written in Python, is available online:
\begin{quote}
 \url{https://github.com/avigad/polya}
\end{quote}
We have named the system ``Polya,'' after George P\'olya, in recognition of his work on inequalities as well as his thoughtful studies of heuristic methods in mathematics (e.g.~\cite{hardy:littlewood:88} \cite{polya:04}).

The general idea of deriving inequalities by putting terms in a normal form and combining specialized modules is found in Avigad and Friedman \cite{avigad:friedman:06}, which examines what happens when the additive and multiplicative fragments of real arithmetic are combined. This is analogous to the situation handled by SMT solvers, with the added twist that the languages in question share inequality symbols and multiplication by constant coefficients in addition to the equality symbol. Avigad and Friedman show that the universal fragment remains decidable even if both theories include multiplication by rational constants, while the full first-order theory is undecidable. The former decidability result, however, is entirely impractical, for reasons discussed there. Rather, it is the general framework for combining decision procedures and the use of canonical normal forms that we make use of here.

The outline of the paper is as follows. In Section~\ref{section:framework}, we describe the general blackboard architecture which is the shared interface for the different modules, and the canonical form for terms. In Section~\ref{section:fourier:motzkin}, we describe the implementation of the additive and multiplicative modules based on the Fourier-Motzkin algorithm, whereas in Section~\ref{section:geometric} we describe the implementation based on existing tools from discrete geometry. In Section~\ref{section:functions}, we describe a module that instantiates general axioms, and in Section~\ref{section:other:modules} we describe more specialized modules that contribute information to the blackboard. In Section~\ref{section:examples}, we provide a number of examples that help characterize the method's strengths and weaknesses. Finally, in Section~\ref{section:conclusions}, we discuss some of the many ways that the method can be extended, as well as ways in which the implementation may be improved.

This paper is a revised and expanded version of the conference paper \cite{avigad:lewis:roux:14}. The extensions described in this paper, chiefly the additional modules described in Section~\ref{section:other:modules}, are due to Avigad and Lewis. More detailed descriptions of some of the representations and algorithms can be found in Lewis' MS thesis \cite{lewis:14}.

\section{The Framework}
\label{section:framework}

\subsection{Terms and Canonical Forms}
\label{subsection:terms}

We wish to consider terms, such as $3 (5x + 3y + 4 x y)^2 f(u + v)^{-1}$, that are built up from variables and rational constants using addition, multiplication, integer powers, and function application. To account for the associativity of addition and multiplication, we view sums and products as multi-arity rather than binary operations. We account for  commutativity by imposing an arbitrary ordering on terms, and ordering the arguments accordingly.

Importantly, we would also like to easily identify the relationship between terms $t$ and $t'$ where $t = c \cdot t'$, for a nonzero rational constant $c$. For example, we would like to keep track of the fact that $4 y + 2 x$ is twice $x + 2 y$. Towards that end, we distinguish between ``terms'' and ``scaled terms'': a scaled term is just an expression of the form $c \cdot t$, where $t$ is a term and $c$ is a rational constant. We refer to ``scaled terms'' as ``s-terms'' for brevity.

\begin{definition}
  We define the set of \emph{terms} ${\cal T}$ and \emph{
    s-terms} ${\cal S}$ by mutual recursion:
\[\def\arraystretch{1.3}
\begin{array}{lcl}
  t, t_i\in {\cal T} & ~:=~ & 1 \mid x \mid \mbox{$\sum_i s_i$} \mid \mbox{$\prod_i t_i^{n_i}$} \mid f(s_1,\ldots, s_n)\\
  s, s_i\in {\cal S} & ~:=~ & c \cdot t\;.
\end{array}
\]
Here $x$ ranges over a set of \emph{variables}, $f$ ranges over a set of \emph{function symbols}, $c \in \QQ$, and $n_i \in \ZZ$.
\end{definition}
Thus we view $3 (5x + 3y + 4 x y)^2 f(u + v)^{-1}$ as an s-term of the form $3 \cdot t$, where $t$ is the product $t_1^2 t_2^{-1}$, $t_1$ is a sum of three s-terms, and $t_2$ is the result of applying $f$ to the single s-term $1 \cdot (u + v)$. 

Note that there is an ambiguity, in that we can also view the coefficient $3$ as the s-term $3 \cdot 1$. This ambiguity will be eliminated when we define a notion of \emph{normal form} for terms. The notion extends to s-terms: an s-term is in normal form when it is of the form $c \cdot t$, where $t$ is a term in normal form. (In the special case where $c = 0$, we require $t$ to be the term $1$.) We also refer to terms in normal form as \emph{canonical}, and similarly for s-terms.

To define the notion of normal form for terms, we fix an ordering $\prec$ on variables and function symbols, and extend that to an ordering on terms and s-terms. For example, we can arbitrarily set the term $1$ to be minimal in the ordering, then variables, then products, then sums, and finally function applications, recursively using lexicographic ordering on the list of arguments (and the function symbol) within the latter three categories. The set of terms in normal form is then defined inductively as follows:
\begin{itemize}
 \item $1, x, y, z, \ldots$ are terms in normal form.
 \item $\sum_{i=1\ldots n} c_i \cdot t_i$ is in normal form provided $c_1 = 1$, each $t_i$ is in normal form, and $t_1 \prec t_2 \prec \ldots \prec t_n$.
 \item $\prod_i t_i^{n_i}$ is in normal form provided each $t_i$ is in normal form, and $1 \neq t_1 \prec t_2 \prec \ldots \prec t_n$.
 \item $f(s_1,\ldots, s_n)$ is in normal form if each $s_i$ is.
\end{itemize}
The details are spelled out in Avigad and Friedman \cite{avigad:friedman:06}. That paper provides an explicit first-order theory, $T$, expressing commutativity and associativity of addition and multiplication, distributivity of constants over sums, and so on, such that the following two properties hold:
\begin{enumerate}
 \item For every term $t$, there is a unique s-term $s$ in canonical form, such that $T$ proves $t = s$.
 \item Two terms $t_1$ and $t_2$ have the same canonical normal form if and only if $T$ proves $t_1 = t_2$.
\end{enumerate}
The results can be straightforwardly extended to terms with arbitrary integer exponents. For example, the term $3 (5x + 3y + 4 x y)^2 f(u + v)^{-1}$ is expressed canonically as $75 \cdot (x + (3/5) \cdot y + (4/5) \cdot x y)^2 f(u + v)^{-1}$, where the constant in the additive term $5x + 3y + 4 x y$ has been factored so that the result is in normal form. 

The semantics we have chosen for expressions $t^n$ when $n$ is negative or zero is that such an expression is assumed to denote a real number, but in case $t$ is $0$ we make no further assumptions about the value of $t^n$. Thus, for example, we do not combine exponents when putting $x^5 x^{-2} x^{-3} x^0$ into canonical form, though $x^2 x^5$ is reduced to $x^7$. We leave it to the multiplicative module to deal with negative exponents appropriately when the base is known to be nonzero.

The two clauses above provide an axiomatic characterization of what it means for terms to have the same canonical form. As discussed in Section~\ref{section:conclusions}, extending the reach of our methods requires extending the notion of a canonical form to include additional common operations.

\subsection{The Blackboard}
\label{subsection:blackboard}

We now turn to the blackboard architecture, which allows modules to share information in a common language. To the addition module, multiplication is a black box; thus it can only make sense of additive information in the shared pool of knowledge. Conversely, the multiplication module cannot make sense of addition. But both modules can make sense of information in the form $t_1 < c \cdot t_2$, where $t_1$ and $t_2$ are subterms occurring in the problem. The blackboard enables modules to communicate facts of this shape. 

When the user asserts a comparison $t > 0$ to the blackboard, $t$ is first put in canonical form, and names $t_0, t_1, t_2, \ldots$ are introduced for each subterm. It is convenient to assume that $t_0$ denotes the canonical term $1$. Given the example in the last section, the method could go on to define
\begin{eqnarray*}
& & t_1 := x, \quad t_2 := y, \quad t_3 := t_1 t_2, \quad 
t_4 := t_1 + (3/5) \cdot t_2 + (4/5) \cdot t_3, \\
& & \quad \quad t_5 := u, \quad t_6 := v, \quad t_7 := t_5 + t_6, \quad t_8 = f(t_7), \quad t_9 := t_4^2 t_8^{-1}
\end{eqnarray*}
In that case, $75 \cdot t_9$ represents $3 (5x + 3y + 4 x y)^2 f(u + v)^{-1}$. Any subterm common to more than one term is represented by the same name. Separating terms in this way ensures that each module can focus on only those definitions that are meaningful to it, and otherwise treat subterms as uninterpreted constants.

Now any comparison $s \bowtie s'$ between canonical s-terms, where $\bowtie$ denotes any of $<, \le, >, \ge, =$, or $\neq$, translates to a comparison $c_i t_i \bowtie c_j t_j$, where $t_i$ and $t_j$ name canonical terms. But this, in turn, can always be expressed in one of the following ways:
\begin{itemize}
 \item $t_i \bowtie 0$ or $t_j \bowtie 0$, or
 \item $t_i \bowtie c \cdot t_j$, where $c \neq 0$ and $i < j$. 
\end{itemize}
The blackboard therefore maintains the following data:
\begin{itemize}
 \item a defining equation for each $t_i$, and
 \item comparisons between named terms, as above.
\end{itemize}
Note that this means that, \emph{a priori}, modules can only look for and report comparisons between terms that have been ``declared'' to the blackboard. This is a central feature of our method: the search is deliberately constrained to focus on a small number of terms of interest. The architecture is flexible enough, however, that modules can heuristically expand that list of terms at any point in the search. For example, our addition and multiplication modules do not consider distributivity of multiplication over addition, beyond multiplication of rational scalars. But if a term $x (y + z)$ appears in the problem, a module could heuristically add the identity $x (y + z) = x y + x z$, adding names for the new terms as needed.

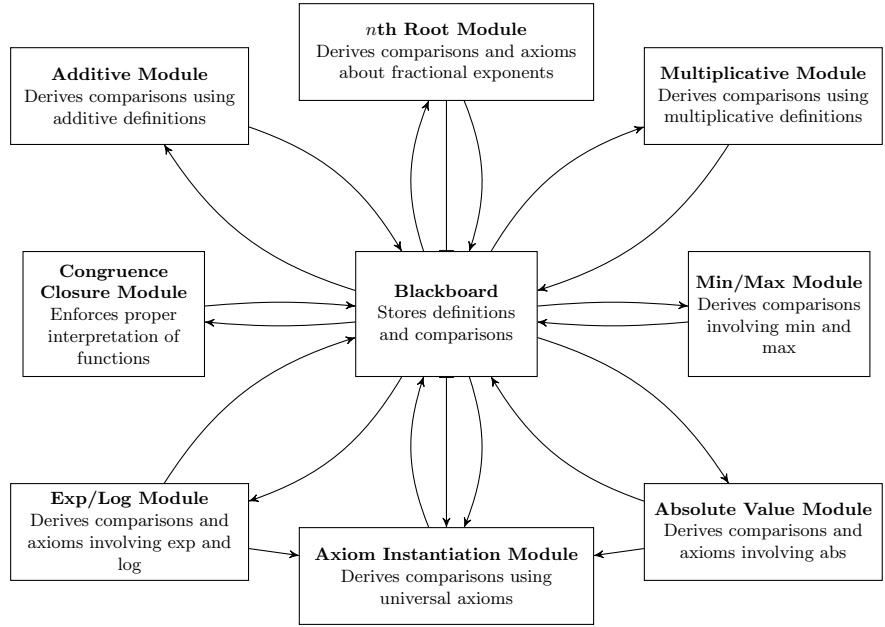
\begin{figure}[h]
\centering
\setlength\fboxsep{3pt}
\setlength\fboxrule{2pt}
{\begin{tikzpicture}[auto, scale=0.5, every node/.style={scale=.75}, node distance=2]
    \node (bb) at (0,2) [rectangle, draw] {\parbox[c][2cm][c]{3cm}{\centering {\bf Blackboard} \\ {\footnotesize Stores definitions and comparisons}} };
  \node (lin) [above left=of bb, rectangle, draw] {\parbox[c][1.5cm][c]{4cm}{\centering {\bf Additive Module} \\ {\footnotesize Derives comparisons using additive definitions}}};
  \node (nonlin) [above right=of bb, rectangle, draw] {\parbox[c][1.5cm][c]{4cm}{\centering {\bf Multiplicative Module} \\ {\footnotesize Derives comparisons using multiplicative definitions}}};
  \node (ax) [below=of bb, rectangle, draw] {\parbox[c][1.5cm][c]{5cm}{\centering {\bf Axiom Instantiation Module} \\ {\footnotesize Derives comparisons using universal axioms}}};
  \node (exp) [below left=of bb, rectangle, draw] {\parbox[c][1.5cm][c]{4cm}{\centering {\bf Exp/Log Module} \\ {\footnotesize Derives comparisons and axioms involving exp and log}}};
  \node (min) [right=of bb, rectangle, draw] {\parbox[c][2cm][c]{3cm}{\centering {\bf Min/Max Module} \\ {\footnotesize Derives comparisons involving min and max}}};
  \node (cc) [left=of bb, rectangle, draw] {\parbox[c][2cm][c]{3cm}{\centering {\bf Congruence Closure Module} \\ {\footnotesize Enforces proper interpretation of functions}}};
  \node (abs) [below right=of bb, rectangle, draw] {\parbox[c][1.5cm][c]{4cm}{\centering {\bf Absolute Value Module} \\ {\footnotesize Derives comparisons and axioms involving abs}}};
  \node (root) [above=of bb, rectangle, draw] {\parbox[c][1.5cm][c]{5cm}{\centering {\bf $n$th Root Module} \\ {\footnotesize Derives comparisons and axioms about fractional exponents}}};
  \draw[->] (bb) [bend left=20] to node {} (lin);
  \draw[->] (bb) [bend left=20] to node {} (nonlin);
  \draw[->] (bb) [bend left=20] to node {} (ax);
  \draw[->] (bb) [bend left=20] to node {} (root);
  \draw[->] (bb) [bend left=20] to node {} (exp);
  \draw[->] (bb) [bend left=5] to node {} (min);
  \draw[->] (bb) [bend left=5] to node {} (cc);
  \draw[->] (bb) [bend left=20] to node {} (abs);
  \draw[->] (lin) [bend left=20] to node {} (bb);
  \draw[->] (nonlin) [bend left=20] to node {} (bb);
  \draw[->] (ax) to [bend left=20] node {} (bb);
  \draw[->] (root) to [bend left=20] node {} (bb);
  \draw[->] (exp) to [bend left=20] node {} (bb);
  \draw[->] (min) to [bend left=5] node {} (bb);
  \draw[->] (cc) to [bend left=5] node {} (bb);
  \draw[->] (abs) to [bend left=20] node {} (bb);

  \draw[->] (abs) to  node {} (ax);
  \draw[->] (exp) to node {} (ax);
  \draw[-|] (root) to node {} (bb);
  \draw[|->] (bb) to node {} (ax);
\end{tikzpicture}}
\caption{The computational structure.}
\label{fig:bbarch}
\end{figure}


To verify an implication, the user asserts the hypotheses to the blackboard, together with the negation of the conclusion. Individual modules then take turns learning new comparisons from the data, and asserting them to the blackboard as well, until a contradiction is obtained, or no further conclusions can be drawn. The setup is illustrated by Figure \ref{fig:bbarch}. Notice that this is essentially the Nelson-Oppen architecture \cite{barrett:et:al:08} \cite{nelson:oppen:79}, in which (disjoint) theories communicate by means of a shared logical symbol, typically equality. Here, the shared language is instead assumed to contain the list of comparisons $<, \le, >, \ge, =, \neq$, and multiplication by rational constants.

Now suppose a module asserts an inequality like $t_3 < 4 t_5$ to the blackboard. It is the task of the central blackboard module to check whether the assertion provides new information, and, if so, to update its database accordingly. The task is not entirely straightforward: for example, the blackboard may already contain the inequality $t_3 < 2 t_5$, but absent sign information on $t_3$ or $t_5$, this does not imply $t_3 < 4 t_5$, nor does the converse hold. However, if the blackboard includes the inequalities $t_3 < 2 t_5$ and $t_3 \leq 7 t_5$, the new assertion is redundant. If, instead, the blackboard includes the inequalities $t_3 < 2 t_5$ and $t_3 \leq 3 t_5$, the new inequality should replace the second of these. A moment's reflection shows that at most two such inequalities need to be stored for each pair $t_i$ and $t_j$ (geometrically, each represents a half-plane through the origin), but comparisons between $t_i$ or $t_j$ and $0$ should be counted among these.

There are additional subtleties: a weak inequality such as $t_3 \leq 4 t_5$ paired with a disequality $t_3 \neq 4 t_5$ results in a strong inequality; a pair of weak inequalities $t_3 \leq 4 t_5$ and $t_3 \geq 4 t_5$ should be replaced by an equality; and, conversely, a new equality can subsume previously known inequalities. The interactions, while not conceptually difficult, are intricate, and care is needed to get the details right.

Below, we will sometimes refer to the terms $t_i$ as the ``problem terms,'' that is, the terms that are registered with the blackboard as objects of comparison.

\subsection{An alternative representation of comparisons}
\label{subsection:alternative:representation}

For each pair of problem terms $t_i$ and $t_j$, we have noted that the blackboard stores the strongest comparison(s) known to hold between them. Sometimes another representation of this information is useful: we can ask for the range of $c$ such that $t_i \leq c t_j$ is known to hold, and the values of $c$ for which the inequality is strict, as well as the dual questions with $\le$ replaced by $\ge$. A moment's reflection shows that if $t_i \leq c_1 t_j$ and $t_i \leq c_2 t_j$, then $t_i \leq c t_j$ for every value $c$ between $c_1$ and $c_2$. Thus, the range of coefficients $c$ for which such a comparison is known form a closed interval, of one of the forms $[a, b]$, $[a, \infty)$, $(-\infty, b]$, or $(-\infty, \infty)$. Moreover, the comparison can be weak or strict at each finite endpoint, as well as weak or strict in the interior. These possibilities are not entirely independent: for example, if the comparison is strict at either endpoint, it will be strict in the interior.

The blackboard's methods are capable of returning such a representation of the comparisons that hold between $t_i$ and $t_j$, in both the $\le$ and $\ge$ directions. More detail can be found in \cite{lewis:14}. This representation is currently used by the minimum module, described in Section~\ref{subsection:minimum}.

\section{Fourier-Motzkin}
\label{section:fourier:motzkin}

The Fourier-Motzkin algorithm \cite{schrijver:86} is a quantifier-elimination procedure for the theory of the structure $\langle \RR, 0, +, < \rangle$, that is, the real numbers as an additive ordered group. Nothing changes essentially if we add to the language of that theory the constant $1$ and scalar multiplication by $c$, for each rational $c$. Here we see that the method can be used to infer comparisons between variables from additive data, and that this can be transported to the multiplicative setting as well.

\subsection{The Fourier-Motzkin Additive Module}
\label{subsection:fm:additive}

The Fourier-Motzkin additive module begins with the comparisons $t_i \bowtie c \cdot t_j$ stored in the blackboard, where $\bowtie$ is one of $\leq, <, \geq, >, =$ (disequalities are not used). It also makes use of comparisons $t_i \bowtie 0$, and all definitions $t_i = \sum_j c_j t_{k_j}$ in which the right-hand side is a sum. The goal is to learn new comparisons of the form $t_i \bowtie c \cdot t_j$ or $t_i \bowtie 0$. The idea is simple: to learn comparisons between $t_i$ and $t_j$, we need only eliminate all the other variables. 
For example, suppose, after substituting equations, we have the following three inequalities:
\begin{eqnarray*}
 3 t_1 + 2 t_2 - t_3 & > & 0 \\
 4 t_1 + t_2 + t_3 & \geq & 0 \\
 2 t_1 - t_2 - 2 t_3 & \geq & 0 
\end{eqnarray*}
Eliminating $t_3$ from the first two equations we obtain $7 t_1 + 3 t_2 > 0$, from which we can conclude $t_1 > (-3 / 7) t_2$. Eliminating $t_3$ from the last two equations we obtain $10 t_1 + t_2 \geq 0$, from which we can conclude $t_1 \geq (-1 / 10) t_2$. More generally, eliminating all the variables other than $t_i$ and $t_j$ gives the projection of the convex region determined by the constraints onto the $i, j$ plane, which determines the strongest comparisons for $t_i$ and $t_j$ that are implied by the data.

Constants can be represented using the special variable $t_0=1$, which
can be treated as any other variable. Thus eliminating all variables except for $t_i$ and $t_0$ yields all comparisons between $t_i$ and a constant.



The additive module simply carries out the elimination for each pair
$i$, $j$. In general, Fourier-Motzkin elimination can
require doubly-exponential time in the number of variables.  With a bit of cleverness, one can use previous eliminations to save some work, but for a problem with $n$ subterms, one is still left with $O(n^2)$-many instances of Fourier-Motzkin with up to $n$ variables in each. 
It is interesting to note that for the examples described in
Section~\ref{section:examples}, the algorithm performs reasonably
well. In Section~\ref{section:geometric}, however, we describe a more
efficient approach.

\subsection{The Fourier-Motzkin Multiplicative Module}
\label{subsection:fm:multiplicative}

The Fourier-Motzkin multiplication module works analogously: given comparisons $t_i \bowtie c \cdot t_j$ or $t_i \bowtie 0$ and definitions of the form $t_i = \prod_j t_{k_j}^{n_j}$, the module aims to learn comparisons of the first two forms. The use of Fourier-Motzkin here is based on the observation that the structure $\langle \RR, 0, +, < \rangle$ is isomorphic to the structure $\langle \RR^+, 1, \times, < \rangle$ under the map $x \mapsto e^x$. With some translation, the usual procedure works to eliminate variables in the multiplicative setting as well. In the multiplicative setting, however, several new issues arise.

First, the multiplicative module only makes use of terms $t_i$ which are known to be strictly positive or strictly negative. The multiplicative module thus executes a preprocessing stage which tries to infer new sign information from the available data. For example, given the definition $t_4 = t_7^3 t_9 t_{11}^2$ and the sign information $t_4 > 0$ and $t_9 < 0$, one can infer $t_7 < 0$ and assert this comparison to the blackboard. The processing phase also infers straightforward inequalities that hold even when sign information is not available; for example, it infers $t_i t_j < t_i t_k$ whenever $t_i > 0$ and $t_j < t_k$ are known, even if the signs of $t_j$ and $t_k$ are not known. This preprocessing somewhat compensates for the module's need for sign information. However, it is not robust; the more systematic way to accommodate this constraint requires case splitting on the signs of variables. Polya is able to do this in limited settings; see the discussion in Section~\ref{section:conclusions}.


Second, the inequalities that are handled by the multiplicative module
are different from those handled by the additive module, in that terms can have a rational coefficient. For example, we may have an inequality $3 t_2^2 t_5 > 1$; here, the multiplicative constant $3$ would correspond to an additive term of $\log 3$ in the additive procedure. This difference makes it difficult to share code between the additive and multiplicative modules, since it prevents the logarithmic transformation from being carried out explicitly. But these rational coefficients are easy to handle in the multiplicative module. 

Finally, the multiplicative elimination may produce information that
cannot be asserted directly to the blackboard, such as a comparison
$t_i^2 < 3 t_j^2$ or $t_i^3 < 2 t_j^2$. In that case, we have to pay
careful attention to the signs of $t_i$ and $t_j$ and their relation
to $\pm 1$ to determine which facts of the form $t_i \bowtie c \cdot t_j$ can be inferred. We compute exact roots of rational numbers when possible, so a comparison $t_i^2 < 9 t_j^2$ translates to $t_i < 3 t_j$ when $t_i$ and $t_j$ are known to be positive. As a last resort, faced with a comparison like $t_i^2 < 2 t_j^2$, we use a rational approximation of $\sqrt 2$ to try to salvage useful information.

\section{Geometric Methods}
\label{section:geometric}

Although the Fourier-Motzkin modules perform reasonably well on small problems, they are unlikely to scale well. The problem is that many of the inequalities that are produced when a single variable is eliminated are redundant, or subsumed by the others. Thus, by the end of the elimination, the algorithm may be left with hundreds or thousands of comparisons of the form $t_i \bowtie c\cdot t_j$, for different values of $c$. Some optimizations are possible, such as using simplex based methods (e.g.~\cite{dutertre:de:moura:06}) to filter out some of the redundancies. In this section, however, we show how methods of computational geometry can be used to address the problem more directly. On many problems in our test suite, performance is roughly the same. But on some problems of moderate complexity (e.g. example \ref{eq:8i} in Section \ref{section:examples}) we have found our implementation of the geometric approach to be much faster than the Fourier-Motzkin approach. The two methods begin to differ noticeably when the number of problem terms is between 15 and 20. 

\subsection{The Geometric Additive Module}
\label{subsection:additive:geometric}
Geometric methods provide an alternative perspective on the task of
eliminating variables. For real variables $t_i$ and constants $c_i$, a linear inequality $c \leq \sum_{i=1}^k c_i
\cdot t_i$ determines a half-space in $\RR^{k+1}$; when $c=0$, as in the
homogenized inequalities in our current problem, the defining
hyperplane of the half-space contains the origin. A set of $n$
homogeneous inequalities determines an unbounded pyramidal polyhedron
in $\RR^k$ with vertex at the origin, called a ``polyhedral cone.'' (Equalities, represented as $(k-1)$-dimensional hyperplanes, simply reduce the dimension of the polyhedron.) The points inside this polyhedron represent solutions to the inequalities. The problem of determining the strongest comparisons between $t_i$ and $t_j$ then reduces to finding extremal ratios of the $i$-th and $j$-th coordinates of points inside the polyhedron.

We use the following well-known theorem of computational geometry (see \cite[Section 1.1]{ziegler:95}):
\begin{theorem}
 A set $C\subseteq \RR^d$ is a finite intersection of closed homogeneous linear halfspaces (an \emph{$\mathcal{H}$-polyhedron}) if and only if it is a finitely generated conical combination of vectors (a \emph{$\mathcal{V}$-polyhedron}).
\end{theorem}

A description of a $\mathcal{V}$-polyhedron is said to be a
\emph{$\mathcal{V}$-representation} of the polyhedron, and similarly
for $\mathcal{H}$-polyhedrons; there are a number of effective methods to convert between representations. 

\begin{figure}
\centering
\begin{subfigure}{.45\textwidth}
  \centering
  \resizebox{.8\linewidth}{.8\linewidth}{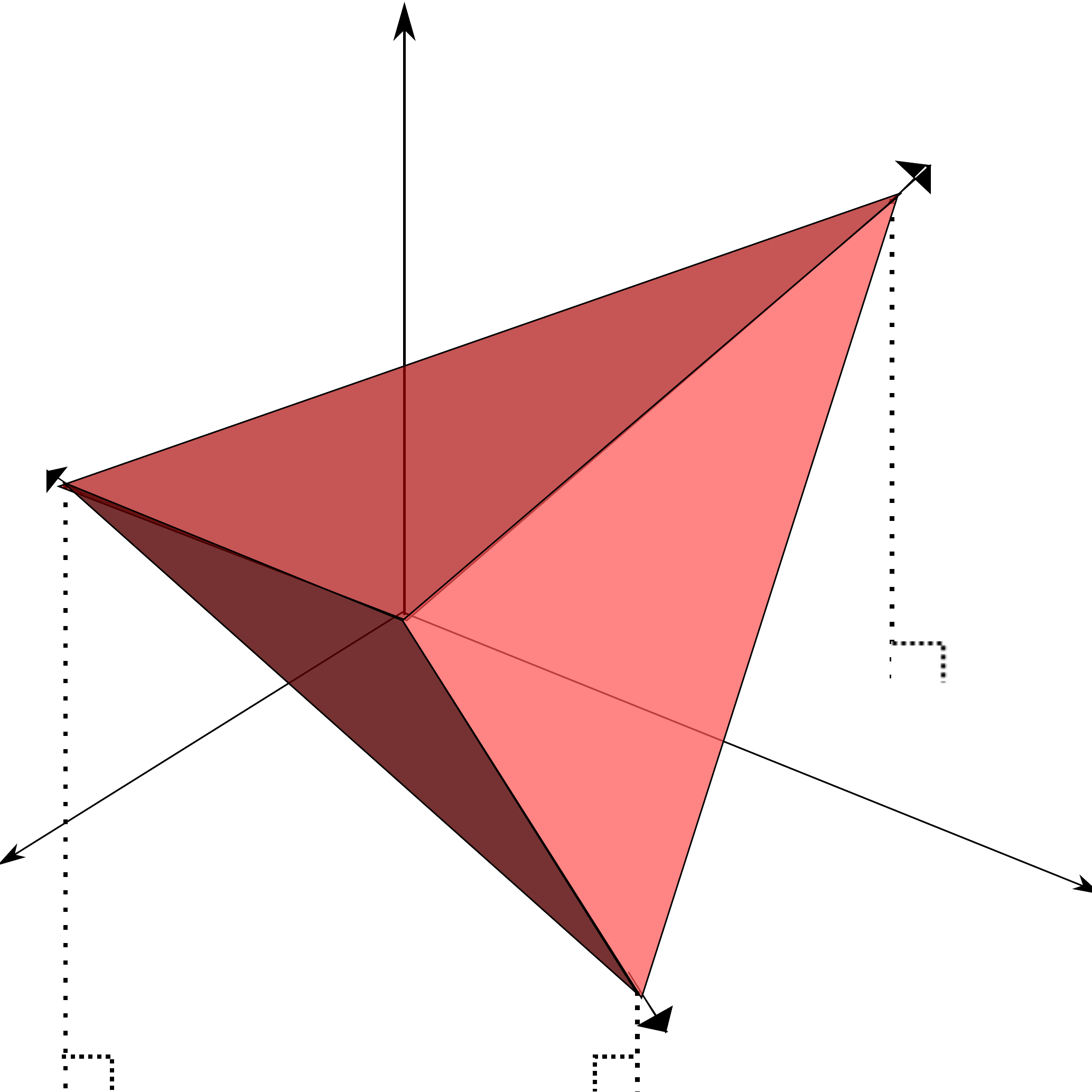}
  \caption{A polyhedral cone in $\RR^3$, defined by three half-spaces}
  \label{fig:geo:projection:a}
\end{subfigure}%
\hspace{.05\textwidth}
\begin{subfigure}{.45\textwidth}
  \centering
  \resizebox{.8\linewidth}{.8\linewidth}{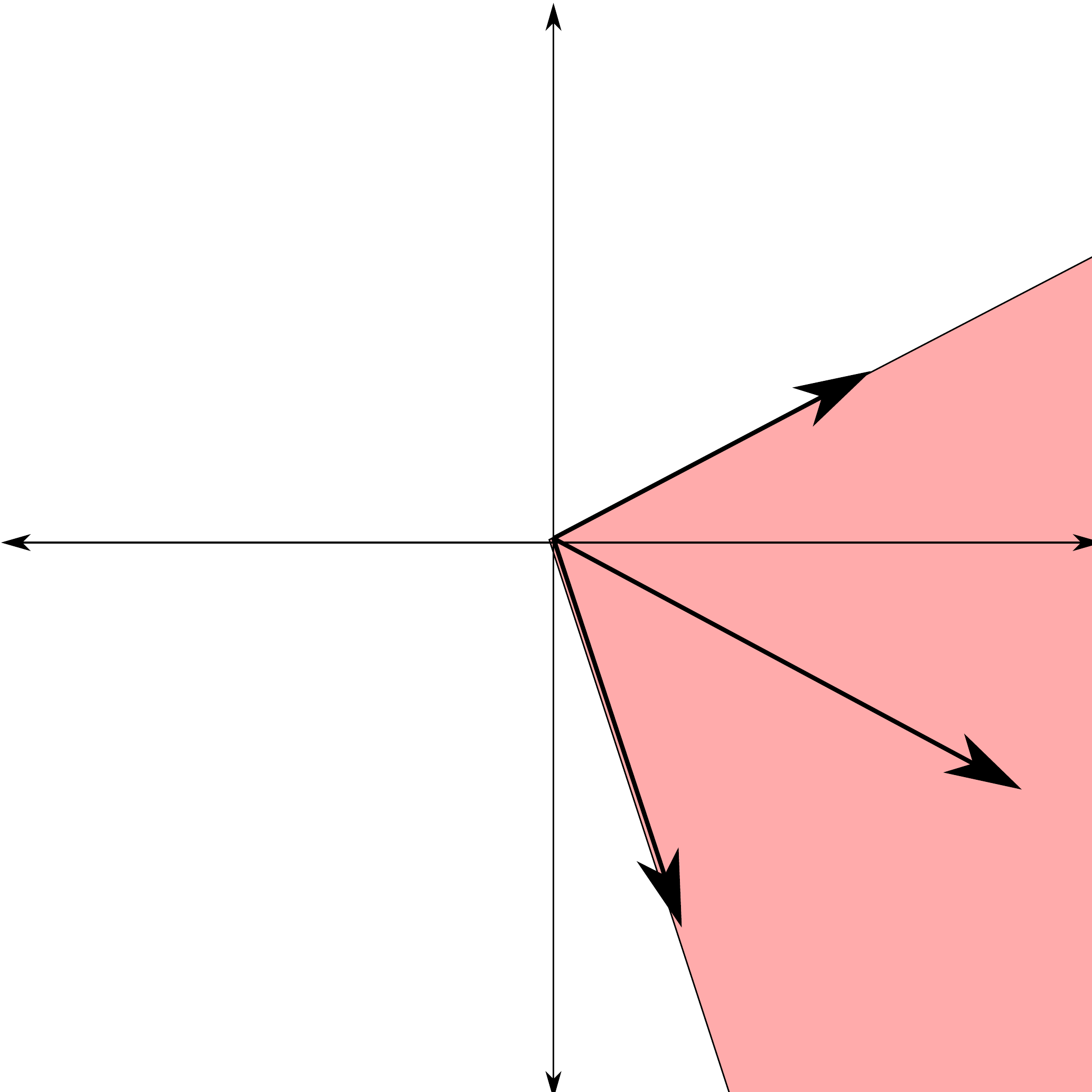}
  \caption{Projected to the $xy$ plane, the polyhedron implies $x\geq2y$ and $x\geq-\frac{1}{3}y$}
  \label{fig:geo:projection:b}
\end{subfigure}
\caption{Variable elimination by geometric projection}
\label{fig:geo:projection}
\end{figure}

The comparisons and additive equalities stored in the central
blackboard essentially describe an $\mathcal{H}$-representation of a
polyhedron. After constructing the corresponding
$\mathcal{V}$-representation, it is easy to pick out the implied
comparisons as follows. For every pair of variables $t_i$ and $t_j$,
project the set of vertices to the $t_i t_j$ plane by setting all the other coordinates to $0$. If there is anything to be learned, all (nonzero) vertices must fall in the same halfplane; find the two outermost points (as in Figure \ref{fig:geo:projection:b}) and compute their slopes to the origin. These slopes determine the coefficients $c$ in two comparisons $t_i \bowtie c \cdot t_j$, and the relative position of the two vertices determine the inequality symbols in place of $\bowtie$.


We chose to use Avis' \emph{lrs} implementation of the reverse-search
algorithm \cite{avis:00} to carry out the geometric computations.
Vertex enumeration algorithms typically assume convexity of the
polyhedron: that is, all inequalities are taken to be weak. As it is
essential for us to distinguish between $>$ and $\geq$, we use a trick
taken from Dutertre and de Moura \cite[Section
5]{dutertre:de:moura:06}. Namely, given a set of strict inequalities $\{0 <
\sum_{i=1}^k c^m_i \cdot t_i: 0\leq m \leq n\}$, we introduce a new
variable $\delta$ with constraints $0 \leq \delta$ and $\{\delta \leq
\sum_{i=1}^k c^m_i \cdot t_i: 0\leq m \leq n\}$, and generate the corresponding polyhedron. The $\delta=0$ hyperplane is assumed to be infeasible.
If, in the vertex representation, every vertex has a zero $\delta$-coordinate, then the inequalities are only satisfiable when $\delta = 0$, which implies that the system with strict inequalities is unsatisfiable. Otherwise, a comparison $t_i \bowtie c\cdot t_j$ is strict if and only if every vertex on the hyperplane $t_i = c\cdot t_j$ has a zero $\delta$ coordinate, and weak otherwise.


\subsection{The Geometric Multiplicative Module}
\label{subsection:multiplicative:geometric}

As with the Fourier-Motzkin method, multiplicative comparisons $1 \leq \prod_{i=1}^k t_i^{e_i}$ can be handled in a similar manner, by restricting to terms with known sign information and taking logarithms. Once again, there is a crucial difference from the additive setting: taking the logarithm of a comparison $c\cdot t_i \cdot t_j^{-1} \bowtie 1$ with $c\neq 1$, one is left with an irrational constant $\log c$, and the standard computational methods for vertex enumerations cannot perform exact computations with these terms.

To handle this situation we introduce new variables to represent the logarithms of the prime numbers occurring in these constant terms. Let $p_1,\ldots, p_l$ represent the prime factors of all constant coefficients in such a problem, and for each $1\leq i \leq l$, let $q_i$ be a variable representing $\log p_i$. We can then rewrite each $c\cdot t_i \cdot t_j^{-1} \bowtie 1$ as $p_1^{d_0}\cdot\ldots\cdot p_l^{d_l} \cdot t_i \cdot t_j^{-1} \bowtie 1$. Taking logarithms of all such inequalities produces a set of additive inequalities in $k+l$ variables. In practice, the factorization problems are small and do not create a bottleneck for our algorithm.

In order to find the strongest comparisons between $t_i$ and $t_j$, we can no longer project to the $t_i t_j$ plane, but instead must look at the $t_it_jq_1\ldots q_l$ hyperplane. The simple arithmetical comparisons to find the two strongest comparisons are no longer applicable; we face the harder problem of converting the vertex representation of a polyhedron to a half-space representation. This problem is dual to the conversion in the opposite direction, and the same computational packages are equipped to solve it. Experimentally, we have found Fukuda's \emph{cdd} implementation of Motzkin's double description method \cite{fukuda:prodon:96} to be faster than \emph{lrs} for this procedure.




\section{The Axiom Module}
\label{section:functions}

The inferences captured by the addition and multiplication modules constitute a fragment of the theory of real-closed fields, roughly, that theory ``minus'' the distributivity of multiplication over addition \cite{avigad:friedman:06}. Recall, however, that we have also included arbitrary function symbols in the language.
An advantage to our framework is that we do not have to treat function
terms as uninterpreted constants; rather, we can seamlessly add
modules that (partially) interpret these symbols and learn relevant inequalities
concerning them.

To start with, a user may wish to add axioms asserting that a particular function $f$ is nonnegative, monotone, or convex. For example, the following axiom expresses that $f$ is nondecreasing:
\[
\forall x, y. \; x \leq y \rightarrow f(x) \leq f(y)
\]
Given such an axiom, Polya's axiom module searches for useful
instantiations during the course of a search, and may thus learn useful information.

Specifically, given a list of universal axioms in variables $v_1\ldots
v_n$, the instantiation module searches for relevant assignments $v_i \mapsto c_i \cdot t_{j_i}$, where each $c_i$ is a constant and each $t_{j_i}$ is a subterm in the given problem. Each axiom is then instantiated with these assignments, and added to the central blackboard as a set of disjunctive clauses. As the search progresses, elements of these clauses are refuted; if only one remains, it is added to the blackboard, as a new piece of information available to all the modules.

The task is a variant of the classic matching problem, but there are at least three aspects of our framework that present complications. First, given that we consider terms with a rational scalar multiplicative constant, the algorithm has to determine those values. So, in the example above, $x$ and $y$ can be instantiated to an s-term $c \cdot t_i$ for any $c$, when such an instantiation provides useful information. Second, we need to take into account the associativity and commutativity of operations like addition and multiplication, so, for example, a term $f(x + y)$ can be unified with a term $f(t_i + t_j + t_k)$ found in the blackboard in multiple ways. Finally, although the framework is built around the idea of restricting attention to subterms occurring in the original problem, at times it is useful to consider new terms. For example, given the axiom
\[
\forall x, y. \; f(x + y) \le f(x) + f(y) \;,
\]
it is clearly a good idea to instantiate $x$ and $y$ to $t_i$ and
$t_j$, respectively, whenever $f(t_i + t_j)$, $f(t_i)$, and $f(t_j)$
all appear in the blackboard, even if the term $f(t_i) + f(t_j)$ does not.

In short, we wish to avoid difficult calculations of rational constants, expensive matching up to associativity and commutativity (see e.g.~\cite{contejean:04}), and unrestrained creation of new terms, while at the same time making use of potentially useful instantiations.
The solution we adopted is to use function terms to trigger and
constrain the matching process, an idea commonly used by SMT solvers \cite{demoura:bjorner:07} \cite{nelson:oppen:79}. Given a universal axiom $(\forall
v_1\ldots v_n)F(v_1,\ldots, v_n)$, $F$ is first converted into clausal
normal form, and each clause $F'$ is treated separately. We take the \emph{trigger set} of $F'$ to be the set of all functional subterms contained in $F$. A straightforward unification procedure finds all assignments that map each trigger element to a (constant multiple of a) problem term, and these assignments are used to instantiate the full clause $F'$. The instantiated clause is asserted to the central blackboard, which checks for satisfied and falsified literals.

For $u$ a term containing unification variables $\{v_i\}$ and $\sigma$ an assignment mapping $v_i\mapsto c_i\cdot t_{j_i}$, the problem of matching $\sigma(u)$ to a problem term $t_{j}$ is nontrivial: the matching must be done modulo equalities stored in the blackboard. For example, if $t_1 = t_2 + t_3$, $t_4 = 2 t_3 - t_5$, and $t_6 = f(t_1 + t_4)$, then given the assignment $\{ v_1 \mapsto t_2 - t_5, v_2 \mapsto 3 t_3 \}$, the term $u = f(v_1 + v_2)$ should be matched to $t_6$. We thus combine a standard unification algorithm, which suggests candidate assignments to the variables occurring in an axiom, with Gaussian elimination over additive and multiplicative equations, to find the relevant matching substitutions.

\section{Additional Modules}
\label{section:other:modules}

In addition to the axiom module, which interprets user-defined functions, Polya incorporates a number of modules which interpret built-in functions. Some of these modules simply assert axioms to the axiom module, and allow it to handle instantiation. Other modules derive information that is too specialized to be handled by the generic axiom module. These modules therefore assert identities, comparisons, and clauses the the blackboard based on special features of the functions and terms they are designed to handle. 

\subsection{The Congruence Closure Module}
Polya's blackboard does not enforce that a function must have the same output given equal inputs. This property is known as \emph{congruence closure}. The well-known union-find data structure and its variations (e.g.\ \cite{demoura:bjorner:07} \cite{Moskal2008}) provides an efficient way to maintain these equalities in a database. This maintenance is not a bottleneck for our algorithm, though, and a more naive approach works well. Polya runs a congruence closure module that searches for pairs of problem terms with the same function symbol and arity. If the Blackboard implies that each corresponding pair of arguments are equal, the module asserts that the terms are themselves equal. The runtime of this module is negligible compared to that of the arithmetical modules, so implementing a more structured method is not a priority.

\subsection{The $n$th Root Module}
Handling terms such as $x^{1/2}$ can be difficult, as this expression is undefined when $x<0$. Canonization must avoid unsound reductions such as simplifying $x^{1/2} \cdot x^{1/2}$ to $x$. On the other hand, when $x > 0$ is known, additional reductions and inferences can be carried out. The canonizer interprets terms $t^{m / n}$ as $(t^{1/n})^m$, where $t^{1/n}$ is, in turn, interpreted as a function term $\nroot_n(t)$, where $n$ is a positive integer constant. Reasoning with these functions can largely be handled by the axiom instantiation module, such that for even $n$, inferences about $\nroot_n(t)$ will be made only if $t$ is known to be positive.

The $n$th root module guarantees that the proper axioms for a given problem have been added to the Blackboard. The module finds a list of $n$ such that $\nroot_n(s)$ appears as some problem term, and axiomatizes the behavior of $\nroot_n(\cdot)$ appropriately for each $n$. If $n$ is even, the axioms
\begin{align*}
 \forall x. \; & x \geq 0 \to (\nroot_n(x))^n = x \\
 \forall x. \; & x \geq 0 \to \nroot_n(x) \geq 0
\end{align*}
are added to the instantiation module. If $n$ is odd, the axiom
$$ \forall x. \; (\nroot_n(x))^n = x $$
is added. These conditional identities provide a sound way of reasoning with fractional exponents.

\subsection{The Exponential and Logarithm Module}
\label{subsection:exponential}
Without computing any exact or approximate values, we can describe the exponential function $\fn{exp}(x)=e^x$ as a positive, strictly increasing function defined on all of $\RR$. The module which interprets this function adds axioms asserting these properties to the axiom instantiation module.

Additionally, the exponential function satisfies the identities 
\begin{align*}
 \fn{exp}(c\cdot x) =&\ \fn{exp}(x)^c \\
 \fn{exp}(x_1 + \ldots + x_n) =&\ \fn{exp}(x_1)\cdot \ldots \cdot \fn{exp}(x_n)
\end{align*}
for scalar $c$. These cannot be axiomatized in a way that the instantiation module will recognize, so the exponential module searches for terms of the appropriate forms and adds equalities as appropriate. Note that this operation is potentially expensive, in that it adds extra terms and multiplicative identities to the blackboard.

The natural logarithm function $\fn{log}(x)$ has axioms and identities dual to the exponential. Since $\fn{log}$ is only defined on the positive reals, these axioms are defined conditionally. That is, the module asserts the axiom
$$
 \forall x, y. \; 0 < x \wedge x < y \to \fn{log}(x) < \fn{log}(y) 
$$
to the instantiation module, and only adds identities when the arguments are known to be positive.

\subsection{The Minimum Module}
\label{subsection:minimum}

The minimum function $\fn{min}(x_1, \ldots, x_k)$ is interpreted in the standard way: it returns the value of one of its arguments $x_i$ such that $x_i \leq x_j$ for all $1\leq j \leq k$. Terms involving $\min$ are canonized in a manner similar to the way that sums are canonized: the arguments are listed according to the underlying term order, and a scalar is brought outside the function so that the first argument is an s-term with coefficient 1. Because the minimum function does not have a fixed arity, it cannot be handled by the general axiom module.

For each problem term $t$ of the form $\fn{min}(c_1 \cdot t_1, \ldots, c_k \cdot t_k)$ in the blackboard $B$, the minimum module asserts that $t \leq c_i \cdot t_i$ for $1\leq i \leq k$. As it is useful to find as much sign information as possible for the multiplicative module, the minimum module also checks for $\bowtie\ \in \{<, \leq, \geq, >\}$ if $c_i \bowtie 0$ for all $i$; if so, it asserts that $t \bowtie 0$ as well.

The module must also account for the fact that $t = \fn{min}(c_1 \cdot t_1, \ldots, c_k \cdot t_k)$ is the \emph{smallest} number less than or equal to all of its arguments. If for some constant $d$ and problem term $s$ we have $s \leq d \cdot c_j\cdot t_j$ for all $1\leq j \leq k$, then we also know that $s \leq d\cdot t$. The minimum module uses the Blackboard's methods for finding implied coefficient ranges  to find, for each problem term $s$, an interval $[a, b]$ for which $b\in [a, b]$ implies $s \leq d \cdot c_j\cdot t_j$ holds for all $1\leq j \leq k$. If such an interval exists, the module determines whether the inequalities are strict at the endpoints, and asserts the relevant information to the blackboard.

\subsection{The Absolute Value Module}
\label{subsection:absolute}
Polya has a specialized module for interpreting the absolute value function. The absolute value of an s-term is canonized by bringing the coefficient outside the absolute value, so that the argument is an s-term with coefficient 1. Basic properties of $\abs$ are handled by asserting the following axioms to the axiom module:

\begin{align*}
 \forall x. \; &(\abs(x) \geq 0) \\
 \forall x. \; &(\abs(x) \leq x) \\
 \forall x. \; &(\abs(x) \geq -x) \\
 \forall x. \; &(x \geq 0 \to \abs(x) = x) \\
 \forall x. \; &(x \leq 0 \to \abs(x) = -x). 
\end{align*}

The axiom module cannot handle the triangle inequality in full generality, and so the absolute value module handles this task on its own. Specifically, the module adds comparisons of the forms
\begin{align*}
 \abs(c_1t_1 + \ldots + c_k t_k) & \leq \abs(c_1 t_1) + \ldots + \abs(c_k t_k) \\
 \abs(c_1t_1 + \ldots + c_k t_k) & \geq \abs(c_j t_j) - \left( \abs(c_1 t_1) + \ldots + \abs(c_k t_k) \right).
\end{align*}
Adding these comparisons indiscriminately would necessitate creating new problem terms $\abs(c_j t_j)$ for each argument not already present in the blackboard, which is not likely to be fruitful. The absolute value module takes a more subtle approach, only learning these comparisons if for each $j$, either $\abs(c_j t_j)$ is already a problem term, or the sign of $t_j$ is known (in which case $\abs(c_j t_j)$ is replaced with $\pm c_j t_j$ as appropriate). This approach does not seem to miss any inferences that the indiscriminate approach would capture, since the comparisons learned will only be useful if something is known about each absolute value. The procedure, however, puts additional stress on the modules for arithmetic: for example, the presence of a term $\abs(t_1 + t_2 + t_3)$ can result in four additional linear inequalities with three terms common to all of them.

\subsection{The Built-in Functions Module}
Sometimes, only minimal information about a function is needed to complete a proof; for instance, it may suffice to know that $\fn{sin}(x) \le 1$ or $\fn{floor}(x) \le x$. Polya's built-in functions module will add simple axioms like this for a variety of common functions. Of course, one could create new modules to interpret any of these functions individually, and add more information than the basic properties used here. The goal of this module is to expand Polya's breadth more so than its depth.

The built-in functions module currently axiomatizes $\fn{sin}$ and $\fn{cos}$ as bounded between $-1$ and $1$, $\fn{tan}$ as equal to $\fn{sin} / \fn{cos}$, and $\fn{floor}(x)$ as bounded between $x - 1$ (strictly) and $x$ (weakly).

\section{Examples}
\label{section:examples}

The current distribution of Polya includes a number of examples that are designed to illustrate the method's strengths, as well as some of its weaknesses. For comparison, we verified a number of these examples in Isabelle, trying to use Isabelle's automated tools as much as possible. These include ``auto,'' an internal tableau theorem prover which also invokes a simplifier and arithmetic reasoning methods, and Sledgehammer  \cite{meng:paulson:09} \cite{blanchette:et:al:11}, which heuristically selects a body of facts from the local context and background library, and exports it to various provers. We also sent some of the inferences directly to the SMT solver Z3 \cite{demoura:bjorner:08}. We report on these results below. We also tried a number of these problems with MetiTarski \cite{akbarpour:paulson:08} and ACL2 \cite{acl2}, which are discussed in Section~\ref{section:conclusions}.

\subsection{Successes}
\label{subsection:successes}

To start with, Polya handles inferences involving linear real inequalities, which are verified automatically by many interactive theorem proving systems. It can also handle purely multiplicative inequalities such as
\begin{equation}
\label{eq:1}
 0 < u < v < 1, \; 2 \leq x \leq y \myRightarrow 2 u^2 x < v y^2,
\end{equation}
which are not often handled automatically. It can solve problems that combine the two, like these:
\begin{align}
\label{eq:2}
x > 1 & \myRightarrow (1 + y^2) x > 1 + y^2 \\
\label{eq:3}
0 < x < 1 & \myRightarrow 1 / (1 - x) > 1 / (1 - x^2) \\
\label{eq:3p5}
0 < u < v, \; 0 < z, \; z + 1 < w & \myRightarrow (u + v + z)^3 < (u + v + w)^5
\end{align}
It also handles inferences that combine such reasoning with axiomatic properties of functions, such as: 
\begin{gather}
\label{eq:4}
(\forall x. \; f(x) \leq 1),\ u < v,\ 0 < w \myRightarrow u + w \cdot f(x) < v + w
\\
\label{eq:5}
(\forall x, y. \; x \leq y \rightarrow f(x) \leq f(y)), \; u < v, \; x < y \myRightarrow u + f(x) < v + f(y) 
\end{gather} 
Isabelle's auto and Sledgehammer fail on all of these but (\ref{eq:4}) and (\ref{eq:5}), which are proved by resolution theorem provers. Sledgehammer can verify more complicated variants of (\ref{eq:4}) and (\ref{eq:5}) by sending them to Z3, but fails on only slightly altered examples, such as:
\begin{gather}
\label{eq:6}
(\forall x. \; f(x) \leq 2), \; u < v, \; 0 < w \myRightarrow u + w \cdot (f(x) - 1) < v + w
\\
\label{eq:7}
\begin{split}
(\forall x, y. \; x \leq y \rightarrow f(x) \leq f(y)), \; u < v, \;
1 < v, \; x \leq y \myRightarrow \\
u + f(x) \leq v^2 + f(y)
\end{split}
\\
\label{eq:8}
\begin{split}
(\forall x, y. \; x \leq y & \rightarrow f(x) \leq f(y)), \; u < v,  \; 1 < w, \; 2 < s,\\
& (w + s) / 3 < v, \; x \leq y \myRightarrow u + f(x) \leq v^2 + f(y)
\end{split} 
\end{gather}
Z3 gets most of these when called directly, but also fails on (\ref{eq:7}) and (\ref{eq:8}). Moreover, when handling nonlinear equations, Z3 ``flattens'' polynomials, which makes a problem like (\ref{eq:3p5}) extremely difficult. It takes Z3 a couple of minutes when the exponents $3$ and $5$ in that problem are replaced by $9$ and $19$, respectively. Polya verifies all of these problems in a fraction of a second, and is insensitive to the exponents in (\ref{eq:3p5}). It is also unfazed if any of the variables above are replaced by more complex terms.

Polya has built-in knowledge about functions such as $\fn{exp}$, $\fn{log}$, $\fn{min}$, $\fn{max}$, $\fn{abs}$, $\fn{ceil}$, and $\fn{floor}$. It verifies examples like these:
\begin{gather}
 \label{eq:8a}
 z > \fn{exp}(x), \; w > \fn{exp}(y) \myRightarrow z^3\cdot w^2 > \fn{exp}(3x + 2y) \\
 \label{eq:8b}
 a > 1, \; b \neq 0, \; c > 0, \; \fn{log}(b^2) > 4, \; \fn{log}(c)>1 \myRightarrow \fn{log}(a\cdot b^2 \cdot c^3) > 7 \\
 \label{eq:8c}
 u > 0, \; v > 0, \; x > 0, \; \fn{log}(x) > 2 u + v \myRightarrow x > 1 \\
 \label{eq:8cc}
 x < y, \; u \leq v \myRightarrow u + \fn{min}(x + 2u, y + 2v) \le x + 3v \\
 \label{eq:8d}
 y > 0 \myRightarrow |3 x + 2 y + 5| < 4 \cdot |x| + 3 y + 6 \\
  \label{eq:8e}
 u > 0, \; v > 1 \myRightarrow \sqrt[3]{u^9v^4} > u^3 v
\end{gather}
\noindent It can also handle examples that combine such functions, such as these:
\begin{gather}
 \label{eq:8f}
 \fn{exp}(\fn{max}(|x|, y)) \geq 1 \\
 \label{eq:8g}
 y > \fn{max}(2, 3x), \; x>0 ,\myRightarrow \fn{exp}(4y - 3x) > \fn{exp}(6) \\
\label{eq:8h}
 y > 0 \myRightarrow \fn{log}(1 + |x| + y) > 0 \\
\label{eq:8i}
 |x| < 3, \; |y| < 2, \; w \leq 0 \myRightarrow |x + 2 y + z| \cdot \fn{exp}(w) < 7 + |z| 
\end{gather}
Z3 fails on (\ref{eq:8a}), (\ref{eq:8b}), (\ref{eq:8d}), and (\ref{eq:8i}), even when the relevant properties of $\fn{exp}$, $\fn{log}$, etc.\ are given as axioms. Given the right properties, however, it succeeds on the others. Similarly, Isabelle's auto tactic does well on problems that can combine rules for common functions with linear arithmetic; it solves (\ref{eq:8cc}), (\ref{eq:8d}), (\ref{eq:8f}), and (\ref{eq:8h}), with the additional information that it should case split on the sign of the terms inside the absolute value on (\ref{eq:8d}). It fails when $\fn{log}(1 + |x| + y)$ is replaced by $\fn{log}(1 + |x| + y^4)$ in (\ref{eq:8h}). Sledgehammer verified (\ref{eq:8c}) using the resolution theorem prover Vampire, but neither auto nor Sledgehammer solves the others.

Polya succeeds examples such as
\begin{equation}
\label{eq:9}
 0 < x < y, \; u < v \myRightarrow 2 u + \fn{exp}(1 + x + x^4) < 2 v + \fn{exp}(1 + y + y^4),
\end{equation}
mentioned in the introduction. Sledgehammer verifies this using resolution, and slightly more complicated examples by calling Z3 with the monotonicity of $\fn{exp}$. Sledgehammer restricts Z3 to linear arithmetic so that it can reconstruct proofs in Isabelle, so to verify (\ref{eq:9}) it provides Z3 with the monotonicity of the power function as well. When called directly on this problem with this same information, however, Z3 resorts to nonlinear mode, and fails.

Sledgehammer fails on an example that arose in connection with a formalization of the Prime Number Theorem, discussed in \cite{avigad:et:al:07}:
\begin{equation}
\label{eq:10}
0 \leq n, \; n < (K / 2)x, \; 0 < C, \; 0 < \varepsilon < 1 \myRightarrow \left(1 + \frac{\varepsilon}{3 (C + 3)}\right) \cdot n < K x
\end{equation}
Z3 verifies it when called directly. Sledgehammer also fails on these \cite{avigad:friedman:06}:
\begin{gather}
\label{eq:11}
0 < x < y \myRightarrow (1+x^2)/(2+y)^{17} < (1+y^2)/(2+x)^{10} \\
\label{eq:12}
\begin{split}
0 < x < y & \myRightarrow (1+x^2)/(2+\fn{exp}(y))\geq  (2+y^2)/(1+\fn{exp}(x))
\end{split}
\end{gather}
Z3 gets (\ref{eq:11}) but not (\ref{eq:12}). Neither Sledgehammer nor Z3 get these:
\begin{align}
\label{eq:13}
(\forall x, y. \; f(x + y) = f(x) f(y)), \; a > 2, \; b > 2 & \myRightarrow f(a + b) > 4 \\
\label{eq:14}
(\forall x, y. \; f(x + y) = f(x) f(y)), \; a + b > 2, \; c + d > 2 & \myRightarrow f(a + b + c + d) > 4
\end{align}
Polya verifies all of the above easily.

The following problem was once raised on the Isabelle mailing list:
\begin{equation}
   x>0, y>0, y<1 \myRightarrow (x+y) > xy
\end{equation}
This inference is verified by Z3 as well as Sledgehammer, but both fail when $x$ and $y$ in the conclusion are replaced by $x^{1500}$ and $y^{1500}$, respectively. Polya is insensitive to the exponent.

Let us consider two examples that have come up in recent Isabelle formalizations \cite{avigad:holzl:serafin:unp}. Billingsley  \cite[page 334]{billingsley:95} shows that if $f$ is any function from a measure space to the real numbers, the set of continuity points of $f$ is Borel. Formalizing the proof involved verifying the following inequality:
\begin{multline}
 i \geq 0, \; |f(y) - f(x)| < 1 / (2 (i + 1)), \\ 
 |f(z) - f(y)| < 1 / (2 (i + 1)) \myRightarrow |f(x) - f(y)| < 1 / (i + 1)
\end{multline}
Sledgehammer and Z3 fail on this, while Polya verifies it easily. 

The second example involves the construction of a sequence $f(m)$ in an interval $(a, b)$ with the property that for every $m > 0$, $f(m) < a + (b - a) / m$. The proof required showing that $f(m)$ approaches $a$ from the right, in the sense that for every $x > a$, $f(m) < x$ for $m$ sufficiently large. A little calculation shows that $m \geq (b - a) / (x - a)$ is sufficient. We can implicitly restrict the domain of $f$ to the integers by considering only arguments $\lceil m \rceil$; thus the required inference is
\begin{multline}
(\forall m. \; m > 0 \rightarrow f(\lceil m \rceil) < a + (b - a) / \lceil m \rceil), \\
a < b, \; x > a, \; m \geq (b - a) / (x - a) \myRightarrow f(\lceil m \rceil) < x\;. 
\end{multline}
Sledgehammer and Z3 do not capture this inference, and the Isabelle formalization was tedious. Polya verifies it immediately.

When restricted to problems involving linear arithmetic and axioms for function symbols, the behavior of Z3 and Polya is similar, although Z3 is much more efficient. As the examples above show, Polya's advantages show up in problems that combine multiplicative properties with either linear arithmetic or axioms involving function symbols. In addition, adding certain axioms to Z3 can cause unexpected interactions: the axioms $\forall x. \; \fn{abs}(x)\geq 0$ and $\forall x. \; \fn{abs}(x) \geq x$ jointly cause Z3 to fail, even on problems that do not involve any absolute values.

For the kinds of problems described in this section, time constraints are not a serious issue. Polya solves a test suite of 81 problems, including the ones discussed here, in about 8.5 seconds on an ordinary desktop (with an Intel i7-3770 4 core CPU at 3.4 GHz), using the polytope packages, the full set of modules, and a set of standard axioms. As noted in Sections~\ref{subsection:exponential} and \ref{subsection:absolute}, however, the exponential and absolute value modules put additional stress on the arithmetic modules. Problem~(\ref{eq:8i}) comes close to the limit of what the Fourier-Motzkin procedures can handle, and Polya takes more than a minute on that problem using those procedures. If we eliminate that problem and two similar ones from the test suite, Polya solves the remainder with the Fourier-Motzkin procedures in about 13.5 seconds. Moreover, instructing Polya not to solve Problem (\ref{eq:9}) without using the exponential module reduces the total to less than 11 seconds. Under the same conditions, Polya solves the test suite in 6 seconds using the polytope packages. We expect that various optimizations and improvements are possible.

In a prior version of this paper \cite{avigad:lewis:roux:14}, our test suite included only 51 problems that could be solved without invoking any of the modules described in Section~\ref{section:other:modules} other than the congruence closure module. Polya solved these in about 2 seconds on an ordinary desktop using the polytope packages, and in about 5.5 seconds using Fourier-Motzkin. 

Test files for Isabelle, Z3, MetiTarski, and ACL2, as well as more precise benchmark results, can be found in the distribution.

\subsection{Performance on KeYmaera Examples}
\label{subsection:keymaerea}

KeYmaera is a verification tool for hybrid systems that combines automated deduction, real-algebraic methods, and computer algebra \cite{platzer:08} \cite{platzer:09}. Among other applications, it has been used to verify control systems for transportation systems. The current version of KeYmaera uses Z3 and Mathematica as a backend for solving the algebraic problems it generates. These algebraic problems are often well-suited for Polya's approach. 

We obtained a collection of 4442 problems generated by KeYmaera. With a 3 second timeout and case splitting disabled, Polya was able to verify the unsatisfiability of 4252 (96\%) in about six minutes. (With case splitting enabled, Polya solves an additional 15 problems, but runs for about ten minutes.) While we were unable to obtain direct comparisons, the experimental results in \cite{platzer:09} report a similar percentage of examples solved by the best available methods.

\subsection{Shortcomings}
\label{subsection:shortcomings}

Of course, Polya fails on wide classes of problems where other methods succeed. It is much less efficient than the best linear solvers, for example, and should not be expected to scale to large industrial problems. 

Polya has other shortcomings. Recall that the multiplicative module only takes advantage of equations where the signs of all terms are known. When called directly, the module fails to make the trivial inference
\begin{equation}
 x > 0, \; y < z \myRightarrow x y < x z\;.
\end{equation}
The preprocessing step described in Section \ref{subsection:fm:multiplicative} enables Polya to prove this, but this preprocessing is not robust, and minor adjustments cause Polya to fail:
\begin{equation}
 x > 0, \; x y z < 0, \; x w > 0 \myRightarrow w > yz
 \label{eq:15}
\end{equation}

The problem just described is easily solved by case splitting on the signs of $y$ and $z$. This is an instance of a general heuristic: it is often useful to split on the signs of problem terms involved in multiplicative terms, when the signs of these terms are not known. There are other situations where a case split can help when Polya is stuck. For example, one can split on comparisons in a binary minimum or maximum: Polya proves $\fn{min}(x, y) + \fn{max}(x, y) = x + y$ from either $x \geq y$ or $x \leq y$, but not outright. Similarly, it is generally useful to split on the sign of an absolute value. We have implemented a mechanism whereby modules can suggest useful case splits for the system to try, as the system carries out nested splits to a user-defined maximum depth. The current implementation is naive and inefficient, however, and needs to be improved (see the discussion in the next section).

Polya's strength comes from the fact that rules and axioms are limited to a small list of ``terms of interest'' stored in the blackboard, allowing modules to contribute information in a flexible way while avoiding combinatorial explosion. But this results in a kind of ``tunnel vision,'' causing Polya to miss inferences that require passage through auxiliary terms. For example, Polya fails to validate
\begin{equation}
\fn{log}(1 + x^2 + \fn{exp}(x)) > x,  
\end{equation}
because it does not consider the intermediate term $\fn{log}(\fn{exp}(x))$. If this term is added to the blackboard, Polya easily infers $1 + x^2 + \fn{exp}(x) > \fn{exp}(x)$, and then $\fn{log}(1 + x^2 + \fn{exp}(x)) > \fn{log}(\fn{exp}(x)) = x$. Users can specify such additional terms to consider when posing a problem. Generally speaking, however, it is not an easy task to determine automatically what terms should heuristically be added to the blackboard, and when.

Another shortcoming, in contrast to methods which begin by flattening polynomials, is that Polya does not, \emph{a priori}, make use of distributivity at all, beyond the distributivity of multiplication by a rational constant over addition. Of course, it is by ignoring distributivity that we make the problem modular and tractable; this ignorance is a basic feature of the system, in some sense. However, this leads to some unfortunate consequences. Any reasonable theorem prover for the theory of real-closed fields can easily establish
\begin{equation}
 x^2 + 2 x + 1 \geq 0,
\end{equation}
which can also be obtained simply by writing the left-hand side as $(x + 1)^2$. But, as pointed out by Avigad and Friedman \cite{avigad:friedman:06}, the method implemented by Polya is, in fact, nonterminating on this example.

\section{Conclusions and Future Work}
\label{section:conclusions}

One advantage of the method described here is that it should not be difficult to generate proof certificates that can be verified independently and used to construct formal derivations within client theorem provers. In fact, with only minor modifications to the code, we have implemented rudimentary proof tracing, taking variable eliminations in the Fourier Motzkin modules as primitive proof steps. For procedures using real-closed fields, this is much more difficult; see \cite{mclaughlin:harrison:05} \cite{harrison:07b}.

We tried a number of our test problems in MetiTarski \cite{akbarpour:paulson:08}, which combines resolution theorem proving with procedures for real-closed fields as well as symbolic approximations to transcendental functions. We found that MetiTarski does well on problems in the language of real-closed fields, but not with axioms for interpreted functions, nor with the examples with $\fn{exp}$. An interesting heuristic method, implemented in ACL2, is described in \cite{hunt:et:al:03}. That method is considerably different from ours; for example, it flattens polynomial terms. Working with ACL2 involves importing ``books'' that not only define the concepts in a given domain, but also configure ACL2's automation to adopt suitable proof strategies. We experimented with ACL2 on some of the problems in our test suite, with what we took to be a reasonable set of imports. (We are grateful to Grant Passmore for guidance here.) In this context, ACL2 solved 21 out of the 39 of our benchmark problems that involve only arithmetic. When it came to problems involving extra function symbols, we found that ACL2 was sensitive to the amount of background information provided; it did well with individual properties, such as the property that a function is monotone, but fared less well with large batteries of facts about log and exp. It seems likely, however, that ACL2 can be made to perform better on our benchmarks with a more finely tuned default. It also seems likely that Polya would benefit by incorporating some of ACL2's heuristics. (The preliminary tests described in this paragraph can be found in the Polya repository.)

We envision numerous extensions to our method. One possibility is to implement more efficient case splitting and conflict-driven clause learning (CDCL) search, as do contemporary SMT solvers. For example, recall that the multiplicative routines only work insofar as the signs of subterms are known. It is often advantageous, therefore, to split on the signs on subterms. The current implementation of Polya can do so naively, but contemporary mechanisms for backtracking assumptions are vastly more efficient. Similarly, making the addition and multiplication modules incremental would streamline this as well.

There are many ways our implementation could be optimized, and, of course, we would gain efficiency by moving from Python to a compiled language like C++. We find it encouraging, however, that even our unoptimized prototype performs well on interesting examples. It seems to us to be more important, therefore, to explore extensions of these methods, and try to capture wider classes of inequalities. This includes reasoning with powers and logarithms to an arbitrary base; reasoning about the integers as a subset of the reals; reasoning about common functions, such as trigonometric functions; and heuristically allowing other natural moves in the search, such as flattening or factoring polynomials, when helpful. We would also like to handle second-order operators like integrals and sums, and interact better with external theorem proving methods.

We emphasize again that this method is not designed to replace conventional methods for proving linear and nonlinear inequalities, which are typically much more powerful and efficient in their intended domains of application. Rather, our method is intended to complement these, capturing natural but heterogeneous patterns of reasoning that would otherwise fall through the cracks. What makes the method so promising is that it is open-ended and extensible. Additional experimentation is needed to determine how well the method scales and where the hard limitations lie. 

\medskip

\noindent {\em Acknowledgment.} We are grateful to Leonardo de Moura and the anonymous referees for helpful corrections, information, and suggestions.


\def\cprime{$'$}

\end{document}